\documentclass[conference]{IEEEtran}\usepackage[T1]{fontenc}
\usepackage[utf8]{inputenc}
\usepackage{amsmath} 
\usepackage{amsfonts} 
\usepackage{amssymb} 
\usepackage{amsthm}
\usepackage{color}
\usepackage{url}
\usepackage{caption}
\usepackage{subcaption}
\usepackage[linesnumbered,ruled,vlined]{algorithm2e}
\usepackage{setspace} 
\usepackage{verbatim}
\usepackage{multirow}
\usepackage{graphicx}
\usepackage{comment}

\SetKwInput{KwInput}{Input}                % Set the Input
\SetKwInput{KwOutput}{Output}              % set the
\SetKwInput{Kwinitialize}{Initialization}              % set the

\theoremstyle{remark}

\usepackage{cite}
\renewcommand{\vec}[1]{{\bf{#1}}} % bold symbols
\newcommand{\vecgreek}[1]{{\boldsymbol{#1}}} % bold symbols for lower case greek letters
\newcommand{\tran}{^{\mbox{\scriptsize T}}}

\newcommand{\ex}[2][3]{^{\backslash #2 }}%% Exclude

\newcommand{\gammab}{\vecgreek{\gamma}}

\newcommand{\mybibliography}{\bibliography{conf_short,jour_short,Bi.bib}}
\begin{document}
\title{Device Detection and Channel Estimation   in  MTC with Correlated Activity Pattern}
%\author{\IEEEauthorblockN{Hamza Djelouat,  Mikko Sillanpää, and Markku Juntti}\IEEEauthorblockA{Centre for Wireless Communications -- Radio Technologies,  FI-90014, University of Oulu, Finland\\e-mail: \{hamza.djelouat,markku.juntti\}@oulu.fi.\\\textbf{Area: B. MIMO Communications and Signal Processing}: \\\textbf{Topic: B.1 Multiuser and Massive MIMO}}
%\thanks{The authors are with Centre for Wireless Communications -- Radio Technologies, FI-90014, University of Oulu, Finland. e-mail: \{hamza.djelouat,markus.leinonen,markku.juntti\}@oulu.fi.\\
%

%\textbf{Area: B. MIMO Communications and Signal Processing}: \\\textbf{Topic: B.1 Multiuser and Massive MIMO}}

\author{
\IEEEauthorblockN{Hamza Djelouat,   and Markku Juntti}
\IEEEauthorblockA{\textit{Centre for Wireless Communications -- Radio Technologies} \\
\textit{University of Oulu}\\
Finland\\
e-mail: \{hamza.djelouat,markku.juntti\}@oulu.fi}
\and
\IEEEauthorblockN{Mikko J. Sillanpää}
\IEEEauthorblockA{\textit{Research Unit of Mathematical Sciences} \\
\textit{University of Oulu}\\
Finland\\
e-mail: mikko.sillanpaa@oulu.fi }
\and
\hspace{5cm}\textbf{Area: B. MIMO Communications and Signal Processing} \\\hspace{5cm}\textbf{Topic: B.1 Multiuser and Massive MIMO}

}

\maketitle

\begin{abstract}
%This paper addresses the problem of joint user identification and channel estimation (JUICE) for grant-free access in massive machine-type communications (mMTC). To provide realistic performance analysis, we consider the JUICE under a spatially correlated fading channel model as that reflects the main characteristics of multiple-input multiple-output channels. We formulate the JUICE as a sparse recovery problem in a multiple measurement vector setup and present a solution based on approximate message passing (AMP) that takes into account the channel spatial correlation structure. Furthermore, we derive theoretical analysis on the performance of AMP in terms of activity detection and channel estimation errors.  

%In the latter, The devices are usually grouped into clusters based on location/ functionality creterion.

%In practical machine-type-communications network, the connected devices are grouped into clusterd  activity pattern of the connected devices is event triggred controlled by event-tra the activity pattern alarm traffic,for which the active MTDs are concentrated nearby the epicenter of alarm events,i.e.,they present a high spatiotemporal correlation. 

% based on even In such mode, the devices are grouped into clusters and the activity pattern of the devices within each cluster is correlated.
%, the us the sporadic activity pattern exhibit both a two level sparsity both an clustered the activity of the connected devices is correlated in time and space. Thus, the traffic pa

This paper provides a solution for the  activity detection and channel estimation problem in grant-free access with correlated device activity patterns.  In particular, we consider a machine-type communications (MTC)  network operating in event-triggered traffic mode, where the devices are distributed over clusters with an activity behaviour that exhibits both intra-cluster and inner-cluster sparsity patterns. Furthermore, 
to model the network's intra-cluster and inner-cluster sparsity, we propose a structured sparsity-inducing spike-and-slab prior which provides a flexible approach to encode
the prior information about the correlated sparse activity pattern. Furthermore, we drive a Bayesian inference scheme based on the expectation propagation (EP) framework
to solve the JUICE problem. Numerical results highlight the significant gains obtained by the proposed structured sparsity-inducing spike-and-slab prior in terms of both user identification accuracy and channel estimation performance. 

%in comparison to state-of-the art sparse recovery algorithms in  terms of channel estimation and activity detection performances for clustered user activity patterns.

%Thus, by leveraging the spike and slab sparse prior, we forma structured  To address this problem, we formulate the JUICE problem as a maximum \emph{a posteriori} probability (MAP) problem with properly chosen priors to incorporate the partial knowledge of the UEs' clustered activity and the unknown covariance matrices. We derive a computationally-efficient algorithm based on alternating direction method of multipliers (ADMM) to solve the MAP problem iteratively via a sequence of closed-form updates. 

\end{abstract}

\begin{IEEEkeywords}
Bayesian inference, grant-free MTC, EP, structured sparsity
\end{IEEEkeywords}

%%%%%%%%%%%%%%%%%
\section{Introduction}
% have  widely been  applied in the design of massive machine-type communications (mMTC) solutions with grant-free access protocols. A major challenge in grant-free access is the joint user identification and channel estimation (JUICE). Subsequently, owing to the sporadic nature of activity pattern of the mMTC devices, namely user equipments (UEs), the JUICE has been widely addressed as a sparse recovery problem and solved using various algorithms such as approximate message passing (AMP), sparse Bayesian learning (SBL), and mixed-norm minimization.

Sparse signal recovery techniques have become prevalent in the development of solutions for machine-type communications (MTC) with grant-free access protocols. One of the main challenges in grant-free access is joint user identification and channel estimation (JUICE). Motivated by the sporadic nature of the activity pattern of the MTC devices, namely user equipments (UEs), JUICE has been approached as a problem of sparse recovery and addressed through several algorithms, including approximate message passing (AMP), sparse Bayesian learning (SBL), and mixed-norm minimization. Most of the prior work on the JUICE  considers MTC networks with a random  UE activity pattern \cite{ke2020compressive,liu2018massive,cheng2020orthogonal,djelouat2021spatial}. This could model, e.g., a scenario where UEs monitor independent random processes and thus activate randomly based on certain application criteria.

This paper makes the following distinction from the prior works: we consider an MTC network where the UEs are clustered in groups around the epicentre of alarm-event, thus, rendering their activity highly correlated.   For instance, this models a network where the UEs form clusters based on their geographical locations and each cluster is associated with a monitoring task. Here, an event could trigger a small subset of UEs belonging to a cluster to activate concurrently, leading to clustered UE activity system-wise. 

This paper addresses the JUICE in MTC under correlated user activity. More precisely, We propose a solution based on a variational Bayesian inference framework that utilizes a  structured spike-and-slab model \cite{andersen2014bayesian}  to account for such correlation of the UEs sparse activity. Moreover, we derived an expectation propagation-based (EP) algorithm \cite{minka2001expectaion} to solve the Bayesian inference problem under the structured activity pattern. Numerical results demonstrate the clear advantages of the proposed solution over state-of-the-art sparse recovery algorithms.

 %%%%%%%%%%%%%%%%%%%%%%%%%%%%%%% 
\section{System Model and Problem Formulation}
\label{sec::system}

We consider a single-cell uplink network consisting of a set $\mathcal{N}$ of $N$ UEs served by a single BS equipped with a uniform linear array (ULA) of $M$ antennas. The UEs are geographically distributed so that they form $N_c$ \textit{clusters}. For simplicity, we assume that each cluster contains $L$ UEs such that ${N=LN_c}$, but the extension to a more general case is conceptually straighforward. %We denote each cluster as $\mathcal{C}_l$, ${l=1,\ldots,C}$. 
%We denote the index set of the $l$th  cluster as $\mathcal{C}_l$, ${l=1,\ldots,C}$. 
A cluster containing a subset of UE indices is denoted by $\mathcal{C}_{l}\subseteq\{1,2,\ldots,N\}$. We consider a block Rayleigh fading channel response $\vec{h}_i \sim(\vec{0},\beta_i\vec{I}_M)\in\mathbb{C}^{M}$, where $\beta_i$ represents the unknown path-loss and shadowing component. In addition, the BS assigns to each UE ${ i\in \mathcal{N}}$  a unique unit-norm pilot sequence $\vecgreek{\phi}_i \in \mathbb{C}^{\tau_{\mathrm{p}}}$.  Accordingly, the received signal associated with the transmitted pilots at the BS, $\vec{Y} \in \mathbb{C}^{\tau_{\mathrm{p}}\times M}$,  is given by
  \begin{equation}
\label{eq::Y}
 \vec{Y}=\sum_{i=1}^{N}\gamma_i \vecgreek{\phi}_i\vec{h}_i\tran+\vec{W}=\vec{\Phi} \vec{X}\tran+ \vec{W},
\end{equation}
where $\gamma_i=0$ when the $i$th is active and $\gamma_i=0$ when $i$th UE is inactive, $\vec{W} \!\!\in\!\mathbb{C}^{\tau_{\mathrm{p}}\times M}$ is an additive white Gaussian noise with independent and identically distributed (i.i.d.) elements as $\mathcal{CN}(0,\,\sigma^{2})$, $\vec{\Phi}=[\vecgreek{\phi}_1,\ldots,\vecgreek{\phi}_N] \in \mathbb{C}^{\tau_\mathrm{p}\times N}$, and $\vec{X}=[\vec{x}_1,\ldots,\vec{x}_{N}] \in \mathbb{C}^{M\times N}$, with $\vec{x}_i=\gamma_i\vec{h}_i$.

In contrast to the majority of the literature on grant-free access MTC that consider random UE activation, we consider herein the following technical observations on MTC under the event-triggered traffic model: i) the UEs activity is triggered by event concentrated around a very small subset of  \emph{active} clusters, thus, giving rise to an \emph{inner cluster} sparsity structure.  ii) An active cluster refers to any cluster with at least one active UE, while containing at most ${L_\mathrm{c}\leq L}$ active UEs, thus, inducing a correlation between the UEs activity in the form of \emph{intra-cluster} sparsity structure. 

Therefore, in order to encode the prior knowledge on both the intra and inner-cluster sparsity of the network, we  introduce first  the following parameters:
\begin{enumerate}
    \item The binary indicator variable $c_l$, $l=1,\ldots,N_c$,  that controls the intra-cluster sparsity,   defined as $c_l=1$ if the  $l$th cluster is active, and  $c_l=0$ otherwise. Thus, we can statistically model $c_l$ as a  Bernoulli random variable with $p(c_l=1) =\epsilon$ and $p(c_l=0) = 1-\epsilon$. 
    \item The hyper-parameter  $\bar{\gamma}_i \in \mathbb{R}^{+}$, $i \in \mathcal{N}$ that controls  model the intra-cluster sparsity. Ideally, we aim to  estimate $\bar{\gamma_i}=\gamma_i\beta_i$.
\end{enumerate}
Subsequently, we can model the effective channel $\vec{x}_i$, $ \mathcal{N}$, using the the structured spike-and-slab prior as
\begin{equation}\label{spike-slab}
     p(\vec{x}_i|c_l,\bar{\gamma_i})=(1-c_l)\delta(\vec{x}_i)+c_l \mathcal{CN}(\vec{x}_i;\vec{0},\bar{\gamma}_i\vec{I}_M).
\end{equation}
The main idea in \eqref{spike-slab} can be summarized as follow
\begin{itemize}
    \item If $c_l=0$, the vector $\vec{x}_i$ would have only the spike component, delta function, from \eqref{spike-slab}, thus estimated as $\vec{x}_i=\vec{0}$.
    \item  If $c_l=1$, $\vec{x}_i$ would have only the slab component from \eqref{spike-slab} in the form be a Gaussian random vector with covariance $\bar{\gamma_i}\vec{I}_M$. Therefore,  if  $\bar{\gamma}_i\approx 0$,  the variance of the  slab component in \eqref{spike-slab}  would be very small  that we could safely estimate that $\vec{x}_i \approx \vec{0}$, whereas if $\bar{\gamma}_i>0$, $\vec{x}_i$ would be a non-zero Gaussian random vector.
\end{itemize}
%where we have introduced the following parameters: 1) the  binary indicator variable $c_l$, $l=1,\ldots,N_c$  that  controls the intra-cluster sparsity,   defined as $c_l=1$ if the  $l$th cluster is active, and  $c_l=0$ otherwise. Thus, we can statistically model $c_l$ as a  Bernoulli random variable with $p(c_l=1) =\epsilon$ and $p(c_l=0) = 1-\epsilon$.  2) The hyper-parameter  $\bar{\gamma}_i \in \mathbb{R}^{+}$, $i \in \mathcal{N}$ that controls  model the intra-cluster sparsity.  
 %, such that $\vec{x}_i \sim \mathcal{CN}(\vec{0}, \bar{\gamma}_i\vec{I}_M)$
%    \begin{equation}
%c_l =
%\begin{cases}
%%0 , & \text{otherwise}, 
%\end{cases} \quad i=l,\ldots,N_c,
%\end{equation}
%a is the UE activity is imposed by the priors presented in Section \ref{sec:p(gamma)}.%%%%%%%%%%%%
\section{A Bayesian Inference solution via EP}
\label{sec::AMP}
The JUICE problem can be formulated from a Bayesian perspective as  maximum \emph{a posteriori} probability (MAP) problem as follows
%\begin{equation}\footnotesize\begin{array}{ll}\label{eq:map_x_gamma}
%\{\hat{\vec{X}},\hat{\vec{c}}\}&\hspace{-3mm}=\underset{\vec{X},\vec{c}}{\max}~\displaystyle p(\vec{X},\vec{c}|\vec{Y})\\&\hspace{-3mm}=\underset{\vec{X},\gamma,\vec{c}}{\max}~\displaystyle \frac{1}{p(\vec{Y})}p(\vec{Y}|\vec{X})p(\vec{X}|\gamma,\vec{c})p(\vec{c})\\&\hspace{-3mm}=\dfrac{1}{p(\vec{Y})}\underbrace{(\mathcal{CN}(\vec{Y}; \vec{\Phi}\vec{X},\sigma^2\vec{I})}_{f_1(\vec{X})}\underbrace{\prod_{l=1}^{C}\big[(1-c_l)\delta\big(\vec{X}_{\mathcal{C}_l}\big)+c_l\prod_{i \in {\mathcal{C}_l}} \mathcal{CN}(\vec{x}_i;\vec{0},\gamma_i\vec{I}_M) \big]}_{f_2(\vec{X},\vec{c})}\\&\underbrace{\prod_{l=1}^C\mathcal{B}(\vec{c}_l|\epsilon)}_{f_3(\vec{c})} \end{array}\end{equation}
\begin{equation}\footnotesize
\begin{array}{ll}\label{eq:map_x_gamma}
\{\hat{\vec{X}},\hat{\vec{c}},\hat{\gammab}\}&\hspace{-3mm}=\underset{\vec{X},\vec{c},\bar{\gammab}}{\max}~\displaystyle p(\vec{X},\vec{c},\hat{\gammab}|\vec{Y})=\underset{\vec{X},\vec{c},\bar{\gammab}}{\max}~\displaystyle \frac{1}{p(\vec{Y})}p(\vec{Y}|\vec{X})p(\vec{X}|\bar{\gammab},\vec{c})p(\vec{c})\\
&\hspace{-3mm}=\underset{\vec{X},\vec{c},\bar{\gammab}}{\max}~\displaystyle\frac{1}{p(\vec{Y})}f_1(\vec{X})f_2(\vec{X},\vec{c},\bar{\gammab})f_3(\vec{c}),
%&\hspace{-3mm}=\dfrac{1}{p(\vec{Y})}\underbrace{(\mathcal{CN}(\vec{Y}; \vec{\Phi}\vec{X},\sigma^2\vec{I})}_{f_1(\vec{X})}\underbrace{\prod_{l=1}^{C}\big[(1-c_l)\delta\big(\vec{X}_{\mathcal{C}_l}\big)+c_l\prod_{i \in {\mathcal{C}_l}} \mathcal{CN}(\vec{x}_i;\vec{0},\gamma_i\vec{I}_M) \big]}_{f_2(\vec{X},\vec{c})}\\&\underbrace{\prod_{l=1}^C\mathcal{B}(\vec{c}_l|\epsilon)}_{f_3(\vec{c})} 
\end{array}
\end{equation}
where 
\begin{equation}\footnotesize
    \begin{array}{ll}
         f_1(\vec{X})&\hspace{-5mm}=p(\vec{Y}|\vec{X})=\mathcal{CN}(\vec{Y}; \vec{\Phi}\vec{X},\sigma^2\vec{I}),  \\
         f_2(\vec{X},\vec{c},\bar{\gammab})&\hspace{-3mm}=p(\vec{X}|\gamma,\vec{c})=\displaystyle\prod_{l=1}^{N_c}  f_2(\vec{X}_{\mathcal{C}_l},c_l,\bar{\gammab}_{\mathcal{C}_l})\\&=\displaystyle\prod_{l=1}^{N_c}\bigg[(1-c_l)\delta\big(\vec{X}_{\mathcal{C}_l}\big)+c_l\prod_{i \in {\mathcal{C}_l}} \mathcal{CN}(\vec{x}_i;\vec{0},\gamma_i\vec{I}_M) \bigg], \\
         f_3(\vec{c})&=p(\vec{c})=\displaystyle\prod_{l=1}^C\mathcal{B}(\vec{c}_l|\epsilon).
    \end{array}
\end{equation}

Unfortunately, the optimization problem \eqref{eq:map_x_gamma}  is intractable for large $N$ due to the presence of the delta function. Thus, we settle for an approximated solution to \eqref{eq:map_x_gamma}. In particular, we invoke the expectation propagation (EP)  framework of \cite{hernandez2015expectation}. 

In  EP, the main objective is to approximate iteratively the probability distributions in the true posterior $p(\vec{X},\vec{c},\bar{\gammab}|\vec{Y})$ by a simpler distribution $Q(\vec{X},\gammab,\vec{c})$ that belongs to an exponential family. More precisely,  EP aims is to approximate the factors $f_1(\cdot)$, $f_2(\cdot)$, $f_3(\cdot)$ by $q_1(\cdot)$, $q_2(\cdot)$, $q_3(\cdot)$, respectively, such that 
%such that the joint variations approximation  $Q(\vec{X},\gammab,\vec{c})$ such that
\begin{equation}
   p(\vec{X},\vec{c},\bar{\gammab}|\vec{Y})\approx Q(\vec{X},\gammab,\vec{c})=\frac{1}{K^{\mathrm{EP}}}q_1(\vec{X})q_2(\vec{X},\vec{c})q_3(\vec{c}).
\end{equation}

In the EP framework, each  factor $q_k(\cdot)$, $k=1,2,3$, of the joint variations approximation $Q(\vec{X},\gammab,\vec{c})$ is obtained by minimizing iteratively the Kullback-Leibler divergence \cite{bishop2006pattern}  as
\begin{equation}
    q_k^{*}=\min_{q_k} \mathrm{KL}\bigg( f_k(\cdot)Q\ex{k}(\cdot)|| q_k(\cdot)Q\ex{k}(\cdot)\bigg)
\end{equation}
where $Q\ex{k}(\cdot)=\frac{Q(\cdot)}{q_k(\cdot)}$.

\section{Numerical Results}

\begin{figure*}[t!]
\begin{subfigure}[b]{0.48\textwidth}
    \includegraphics[width=\linewidth]{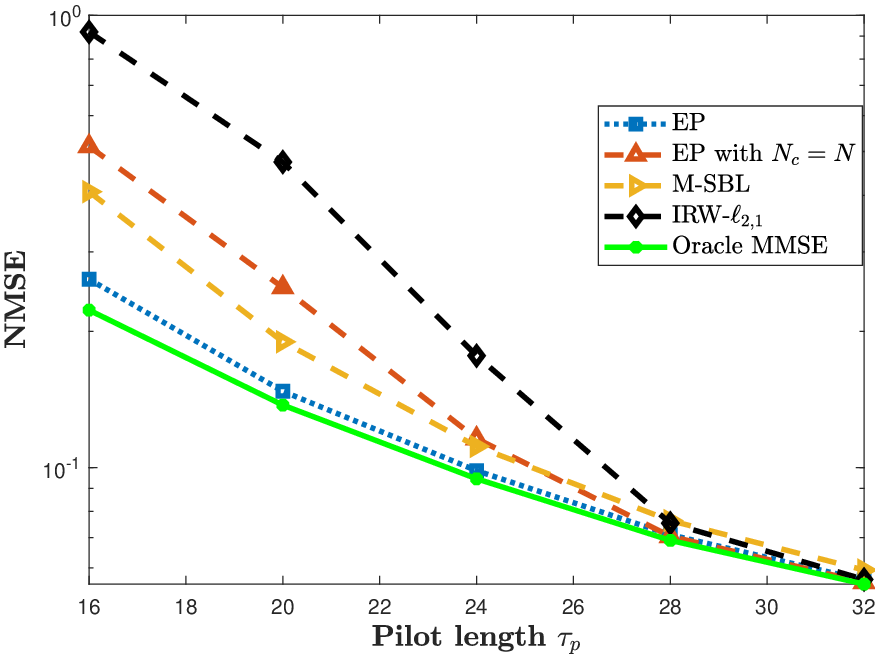}
    \caption{}
    \label{fig:srr_rand}
\end{subfigure}
   \hfill
    \begin{subfigure}[b]{0.48\textwidth}
  \includegraphics[width=\linewidth]{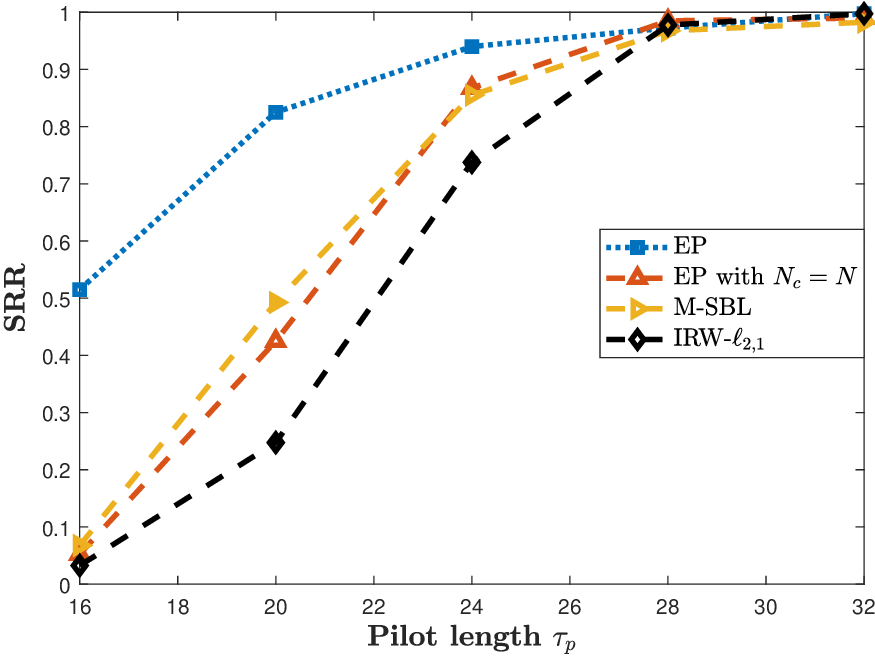}
     \caption{}
     \label{fig:mse_rand}
\end{subfigure}\vspace{-3mm}
\caption{ Performance evaluation of the proposed algorithm with 2 active clusters each containing 8 active UEs, $N=200$, $N_C=20$, $M=10$. }
\label{fig:results_rand}
\vspace{-6mm}
\end{figure*}

Fig.\ \ref{fig:results_rand} compares the performance of \textit{the proposed EP solution } in terms of normalized mean square error (NMSE) and support recovery rate (SRR) against two sparse recovery algorithms: iterative reweighted $\ell_{2,1}$-norm minimization (IRW-$\ell_{2,1}$) \cite{djelouat2021spatial}, and M-SBL \cite{wipf2007empirical} as well as an oracle minimum mean square error (MMSE) estimator that is given both the set of true active UEs and the exact values of $\beta_i$, ${i=1,\ldots,N}$. 

Fig.\ \ref{fig:results_rand} shows two main features for the proposed solution. 1) The proposed algorithm, which considers the activity correlation,  provides a significant gain over M-SBL and IRW-$\ell_{2,1}$ in terms of both channel estimation quality and activity detection accuracy. In fact, the proposed EP algorithm provides near-optimal NMSE performance by approaching the performance provided by the oracle MMSE denoiser which is computed with the set of true active UEs given by the oracle. 2) Although the proposed algorithm performance degrades when the activity correlation is not taken into consideration by setting each cluster to contain only one UE, i.e.,\ ${N_c=N}$, it still outperforms IRW-$\ell_{2,1}$ and matches the performance of M-SBL. 
The obtained results highlight clearly: 1) the importance of using the structured spike-and-slab prior, 2) the gains obtained by using the EP framework to solve the MAP problem.
 
 %the gain obtained by using rich side information at the BS: AMP uses both the channel and the noise statistics, MAP-ADMM uses only the channel statistics while IRW-ADMM operates only on the mere fact that the effective channel is sparse.

\section{Conclusions and Extensions in the Final Paper}
We provided a solution for activity detection and channel estimation in 
grant-free  MTC under correlated activity patterns. First, we introduced the structured spike-and-slab model, which allows for incorporating the prior knowledge of the network traffic pattern. Second, we derived an EP-based approximation to solve the JUICE formulation under the variational Bayesian framework.

\textit{In the final paper, we will provide in detail}  the derivations for the proposed EP algorithm. Furthermore, we will discuss in more detail the computational complexity of the algorithms and propose a few modifications aiming to reduce the computational costs while maintaining the same performance. Finally, we will provide more simulation results to quantify the effect of system parameters, such as the number of BS antennas, transmission power, etc.

%r we analyse the performance in the asymptotic regime, i.e., $N,K,\tau_{\mathrm{p}} \longrightarrow \infty$, with finite ratios  $\frac{N}{\tau_{\mathrm{p}}}$ and $\frac{N}{\tau_{\mathrm{p}}}$  

%\label{conclusion}

\bibliographystyle{IEEEtran} 
\mybibliography 
\end{document}